\documentclass[preprint,showpacs,amsmath,amssymb]{revtex4}
\usepackage{graphicx}
\usepackage{dcolumn}
\usepackage{bm}
\usepackage{psfig}

\newcommand{\BR}{{\cal B}}
\newcommand{\eff}{\varepsilon}

\newcommand{\psip}{\psi(2S)}
\newcommand{\psp}{\psi(2S)}
\newcommand{\jpsi}{J/\psi}

\newcommand{\chicJ}{\chi_{cJ}}
\newcommand{\chicz}{\chi_{c0}}
\newcommand{\chico}{\chi_{c1}}
\newcommand{\chict}{\chi_{c2}}
\newcommand{\EE}{e^+e^-}
\newcommand{\MM}{\mu^+\mu^-}

\newcommand{\pim}{\pi^-}
\newcommand{\piz}{\pi^0}
\newcommand{\pp}{\pi^+\pi^-}
\newcommand{\ppb}{p\overline{p}}

\newcommand{\ppjpsi}{\pi^+\pi^-J/\psi}
\newcommand{\jpsipp}{\pi^+\pi^-J/\psi}
\newcommand{\ra}{\rightarrow}
\newcommand{\jpsito}{J/\psi \rightarrow }
\newcommand{\psipto}{\psi(2S) \rightarrow }
\newcommand{\pspto}{\psi(2S) \rightarrow }

\newcommand{\chicJto}{\chi_{cJ} \rightarrow }
\newcommand{\chiczto}{\chi_{c0} \rightarrow }
\newcommand{\chicoto}{\chi_{c1} \rightarrow }
\newcommand{\chictto}{\chi_{c2} \rightarrow }
\newcommand{\bfg}{\begin{figure}}
\newcommand{\efg}{\end{figure}}
\newcommand{\bitm}{\begin{itemize}}
\newcommand{\eitm}{\end{itemize}}
\newcommand{\bnum}{\begin{enumerate}}
\newcommand{\enum}{\end{enumerate}}
\newcommand{\btbl}{\begin{table}}
\newcommand{\etbl}{\end{table}}
\newcommand{\btbu}{\begin{tabular}}
\newcommand{\etbu}{\end{tabular}}
\newcommand{\aab}{\Lambda\overline{\Lambda}}

\newcommand{\beq}{\begin{equation}}
\newcommand{\edq}{\end{equation}}
\begin{document}

\preprint{Draft-PRD}

\title{\boldmath Determination of $\BR (\chi_{cJ}\rightarrow p \bar{p})$ 
in $\psp$ decays}
\author{
J.~Z.~Bai$^1$,        Y.~Ban$^{10}$,         J.~G.~Bian$^1$,
X.~Cai$^{1}$,         J.~F.~Chang$^1$,       H.~F.~Chen$^{16}$,    
H.~S.~Chen$^1$,       H.~X.~Chen$^{1}$,      J.~Chen$^{1}$,        
J.~C.~Chen$^1$
% ...
\\(BES Collaboration)\\ 
$^1$ Institute of High Energy Physics, Beijing 100039, People's Republic of
     China\\
$^2$ China Center of Advanced Science and Technology, Beijing 100080,
     People's Republic of China\\
$^3$ Guangxi Normal University, Guilin 541004, People's Republic of China\\
$^4$ Henan Normal University, Xinxiang 453002, People's Republic of China\\
$^5$ Huazhong Normal University, Wuhan 430079, People's Republic of China\\
$^6$ Hunan University, Changsha 410082, People's Republic of China\\                                                  
$^7$ Liaoning University, Shenyang 110036, People's Republic of China\\
$^{8}$ Nanjing Normal University, Nanjing 210097, People's Republic of China\\
$^{9}$ Nankai University, Tianjin 300071, People's Republic of China\\
$^{10}$ Peking University, Beijing 100871, People's Republic of China\\
$^{11}$ Shandong University, Jinan 250100, People's Republic of China\\
$^{12}$ Shanghai Jiaotong University, Shanghai 200030, 
        People's Republic of China\\
$^{13}$ Sichuan University, Chengdu 610064,
        People's Republic of China\\                                    
$^{14}$ Tsinghua University, Beijing 100084, 
        People's Republic of China\\
$^{15}$ University of Hawaii, Honolulu, Hawaii 96822\\
$^{16}$ University of Science and Technology of China, Hefei 230026,
        People's Republic of China\\
$^{17}$ Wuhan University, Wuhan 430072, People's Republic of China\\
$^{18}$ Zhejiang University, Hangzhou 310028, People's Republic of China
}
\date{\today}

\begin{abstract}

The processes $\psipto \gamma \chicJ$, $\chicJto \ppb$ ($J=0,1,2$) are
studied using a sample of $14 \times 10^6$ $\psip$ decays collected
with the Beijing Spectrometer at the Beijing Electron-Positron
Collider. Very clear $\chicz$, $\chico$ and $\chict$ signals
are observed, and the branching fractions $ \BR(\chicJto \ppb$)
($J=0,1,2$) are determined to be $ (27.1^{+4.3}_{-3.9}\pm4.7)\times
10^{-5}$, $ (5.7^{+1.7}_{-1.5}\pm0.9)\times 10^{-5}$, and $
(6.5^{+2.4}_{-2.1}\pm1.0)\times 10^{-5}$, respectively, where the first
errors are statistical and the second are systematic.

\end{abstract}

\maketitle

\section{Introduction}

Hadronic decay rates of P-wave quarkonium states provide good tests of
quantum chromodynamics (QCD). The decays $\chicJto \ppb$ have been
calculated using different models~\cite{besi4,besi5}, and recently, the
decay branching fractions of $\chicJto$ baryon and anti-baryon pairs were
calculated including the contribution of the color-octet fock (COM)
states~\cite{wong}.  Using the $\chicJto \ppb$ branching fractions as
input to determine the matrix element, the partial widths of $\chicJto
\aab$ are predicted to be about half of those of $\chicJto \ppb$, for
$J=1$ and $2$.  However, recent measurements of $\chicJto
\aab$~\cite{aa} together with the branching fractions of $\chicJto \ppb$
from $\ppb$ annihilation experiments~\cite{e760,e8351,e8352} and from a
measurement from $\psp$ decays~\cite{besi} seem to
contradict this prediction.  An improved measurement of the $\chicJto
\ppb$ branching fraction with the same data sample that was used for
$\chicJto \aab$ measurement will yield a more consistent measurement
of this fraction.

The measurements of $\BR(\chicJto \ppb)$ have been performed in $\EE$
collision experiments, where the $\chicJ$ are produced in $\psip$
radiative decays, and in $\ppb$ annihilation experiments, where $\chicJ$
are formed directly.  Although the precision in these experiments is
limited, results from $\ppb$ annihilation experiments seem
systematically higher than those obtained in $\EE \ra \psip$
experiments, as shown in Table~\ref{comp}. This led to a global fit
based on results from both $\EE$ and $\ppb$
annihilations~\cite{patrignani}.

\begin{table}[htbp]
\caption{\label{comp} $\BR(\chicJto \ppb)$ results obtained by
    different experiments. } 
\begin{center}
\begin{tabular}{c|c|c}\hline\hline
 Channel &\multicolumn{2}{c}{Experimental technique}\\\cline{2-3}
 &$\psipto \gamma \chicJto \gamma \ppb$ & $\ppb \ra \chicJto \gamma \jpsi$\\
\hline
 $\BR(\chiczto \ppb)(\times 10^{-5})$  & 15.9$\pm$4.3$\pm$5.3~\cite{besi}
                                        &$48^{+9+21}_{-8-11}$~\cite{e8351}   \\
 &                                     &$41 \pm 3^{+16}_{-9}$~\cite{e8352} \\
 $\BR(\chicoto \ppb)(\times 10^{-5})$  &  4.2$\pm$2.2$\pm$2.8~\cite{besi}
                                        &7.8$\pm$1.0$\pm$1.1~\cite{e760} \\
 $\BR(\chictto \ppb)(\times 10^{-5})$  &  5.8$\pm$3.1$\pm$3.2~\cite{besi}
                                        &9.1$\pm$0.8$\pm$1.4~\cite{e760}\\
\hline
\end{tabular}
\end{center}
\end{table}

The above $\EE$ annihilation results \cite{besi} were obtained from a
sample of $3.79 \times 10^6$ $\psip$ events collected with the Beijing
Spectrometer (BESI) detector~\cite{bes} at the Beijing
Electron-positron Collider (BEPC) storage ring running at the energy
of the $\psip$.  Here we report on a result obtained with a sample of
$(14 \pm 0.6)\times 10^6$ $\psip$ events collected with the upgraded
BESII detector~\cite{bes2}. In BESII, a 12-layer vertex chamber (VTC)
surrounding the beam pipe provides trigger information. A forty-layer
main drift chamber (MDC), located radially outside the VTC, provides
trajectory and energy loss ($dE/dx$) information for charged tracks
over $85\%$ of the total solid angle.  The momentum resolution is
$\sigma _p/p = 0.018 \sqrt{1+p^2}$ ($p$ in $\hbox{\rm GeV}/c$), and
the $dE/dx$ resolution for hadron tracks is $\sim 8\%$.  An array of
48 scintillation counters surrounding the MDC measures the
time-of-flight (TOF) of charged tracks with a resolution of $\sim 200$
ps for hadrons.  Radially outside the TOF system is a 12 radiation
length, lead-gas barrel shower counter (BSC).  This measures the
energies of electrons and photons over $\sim 80\%$ of the total solid
angle with an energy resolution of $\sigma_E/E=21\%/\sqrt{E}$ ($E$ in
GeV).  Outside of the solenoidal coil, which provides a 0.4~Tesla
magnetic field over the tracking volume, is an iron flux return that
is instrumented with three double layers of counters that identify
muons of momentum greater than 0.5~GeV/$c$.

A Monte Carlo~(MC) simulation is used for the determination of mass
resolution and detection efficiency. The angular distribution of
$\psipto \gamma X$ is simulated assuming a pure E1 transition, namely
$1+\cos^2\theta$, $1-\frac{1}{3}\cos^2\theta$, and
$1+\frac{1}{13}\cos^2\theta$ for the spin-parity of the resonance $X$
being $J^{P}=0^{+}, 1^{+}$ and $2^{+}$, respectively, and $X$ decaying to
$p \bar{p}$ is simulated according to phase space.
A Geant3 based package, SIMBES, is used for the simulation of detector response,
where the interactions of the produced particles with
the detector material are simulated and detailed consideration of 
the detector performance~(such as dead electronic channels) is included. 
Reasonable agreement between 
data and Monte Carlo simulation has been observed in various 
channels tested, including Bhabha, $e^+ e^- \ra \mu^+ \mu^-$, $\jpsito \ppb$, and 
$\pspto \jpsipp, \jpsito \ell^+\ell^-$ ($\ell=e$ or $\mu$).

\section{Event Selection}

To select $\psipto \gamma \chicJ$, $\chicJto \ppb$ ($J=0,1,2$)
candidates, events with at least one photon and two charged tracks
are required.  
A neutral cluster in the BSC is considered to be a photon candidate when the
angle between the nearest charged track and the cluster 
in the $xy$ plane is greater than $15^{\circ}$, the first
cell hit is in the beginning 6 radiation lengths, and 
the angle between the cluster development direction in the BSC 
and the photon emission direction in $xy$ plane is less than 
$37^{\circ}$. 

A likelihood method is used for discriminating pion, kaon, proton, and
antiproton tracks. For each charged track, an estimator is defined as
$ W^{i}=\frac{P^{i}}{\sum\limits_{i} P^{i}}$, $P^{i}=\prod_{j}
P_{j}^{i}(x_{j})$, where $P^{i}$ is the probability under the
hypothesis of being type $i$, $i=\pi$, $K$, and $p$ or $\bar{p}$, and
$P_{j}^{i}(x_{j})$ is the probability density for the hypothesis of
type $i$, associated with the discriminating variable
$x_{j}$. Discriminating variables used for each charged track are
the time of flight in the TOF and the energy loss of the track in the
MDC.  By definition, pion, kaon, proton and antiproton tracks have
corresponding $W^{i}$ value near one.

For the decay channel of interest, the candidate events are 
required to satisfy the following selection criteria:
\bnum
\item There are two oppositely charged tracks in the MDC with each
  track having a good helix fit and 
$|\cos\theta|<0.75$, where $\theta$ is the polar angle of the track;
\item At least one charged track is identified either as a proton or an anti-proton with
      $W^p>0.7$ or $W^{\bar{p}}>0.7$;
\item There is at least one photon candidate. 
In the case of multiple photon candidates, the one with the largest 
BSC energy is chosen as the  photon radiated from the $\psip$;
\item The $\chi^2$ probability of the four-constraint kinematic fit 
is required to be greater than 1\%.
\enum

\begin{figure}[htbp]
\centerline{\hbox{
\psfig{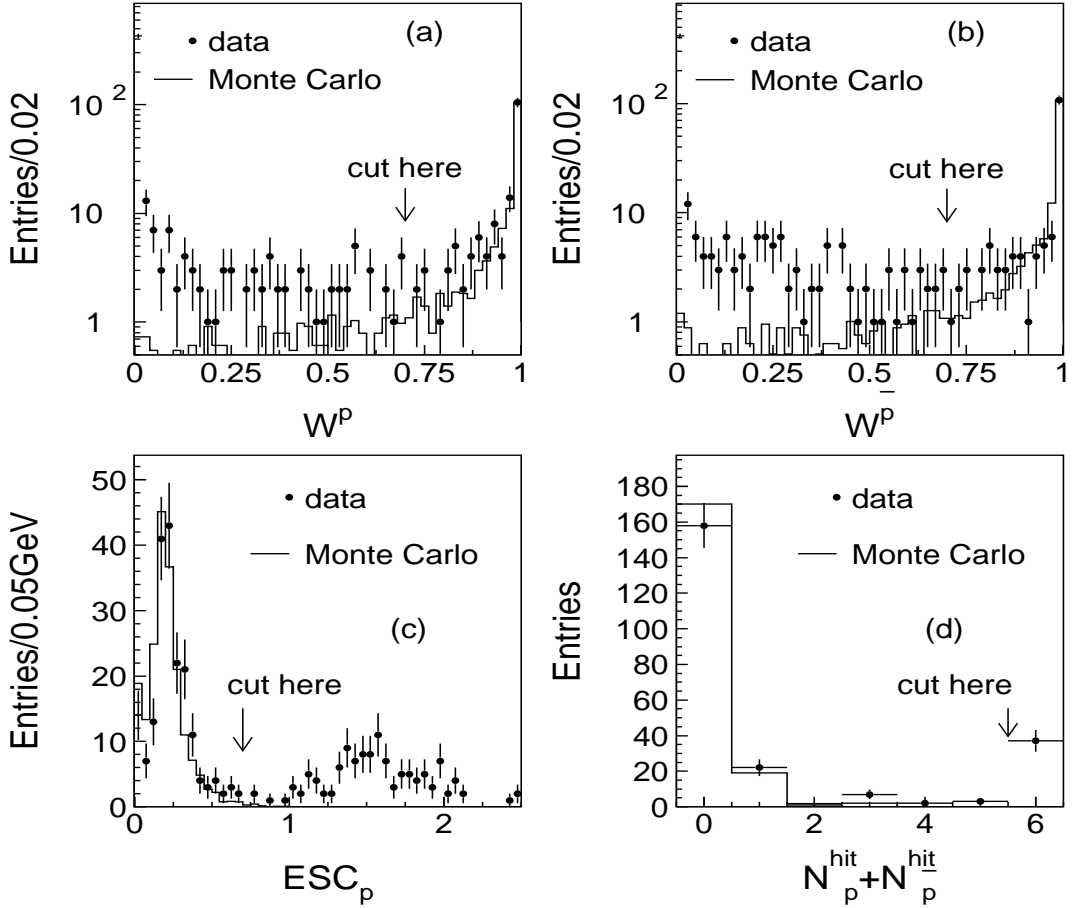}}}
\caption{(a) $W^p$,(b) $W^{\bar{p}}$,(c) $ESC_p$, 
and  (d) $N^{hit}_p+N^{hit}_{\bar{p}}$ distributions after
all the other requirements have been applied, as described in the text. 
The histograms
are for Monte Carlo simulated $\psipto \gamma \chicz$, $\chiczto \ppb$ events
and dots with error bars are for data with $\ppb$ invariant mass
within the
$\chicJ$ signal region. 
For (a) and (b), data and MC simulation are normalized to the last bin, and
for (c) and (d), data and MC simulation are normalized to $ESC_p<0.7$ and 
$N^{hit}_p+N^{hit}_{\bar{p}}<6$, respectively. }
\label{edis}
\end{figure}

Figures~\ref{edis}a and~\ref{edis}b
show distributions of $W^p$ and $W^{\bar{p}}$ after
all other requirements have been applied. It can be seen
that the  $W^{p/\bar{p}}>0.7$ requirement rejects most of the
$\pi$ and $K$ background, while retaining high efficiency.

In order to reduce backgrounds from $\EE$ or $\MM$ tracks in $\jpsi$
decays ($\psipto X \jpsi$), the BSC energy of the positively charged
track~($ESC_p$) is required to be less than 0.7 GeV, and the two charged
tracks must satisfy the muon veto requirement, 
that $N^{hit}_p+N^{hit}_{\bar{p}}<6$, 
where $N^{hit}$ is the number of mu-counter layers with matched hits
and ranges from 0 to 3, indicating not a muon (0), a weak (1), medium
(2), or strongly (3) identified muon track~\cite{muid}.
Distributions of $ESC_p$
and $N^{hit}_p+N^{hit}_{\bar{p}}$ are shown in Figures~\ref{edis}c
  and~\ref{edis}d.

\begin{figure}[htbp]
\centerline{\hbox{
\psfig{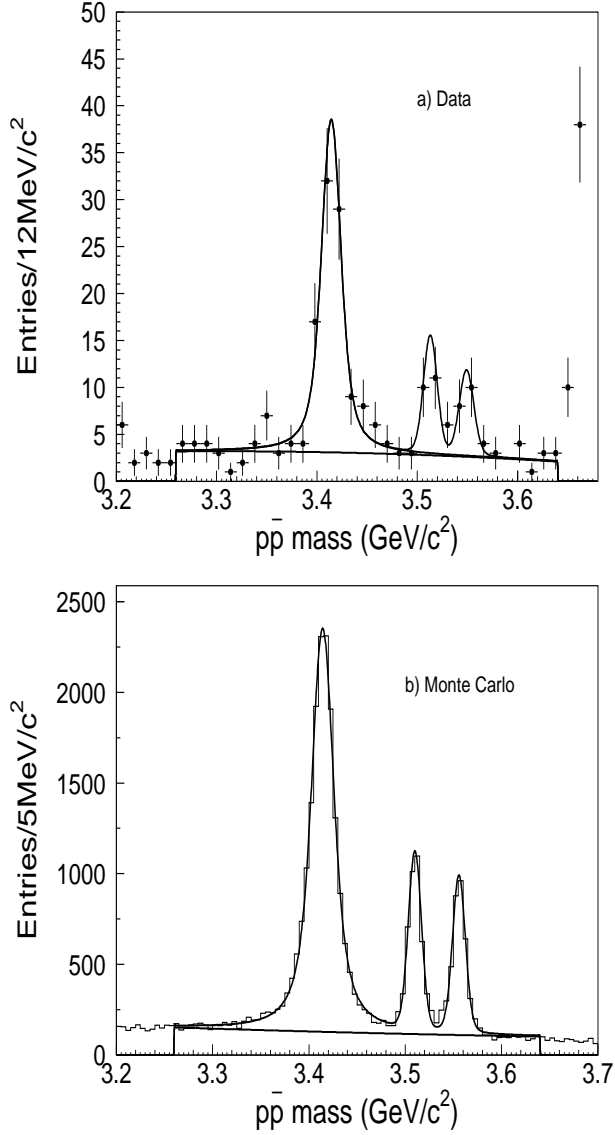}}}
\caption{ Breit-Wigner fit to $p\bar{p}$ mass distribution for selected
$\psi(2S)\rightarrow \gamma p \bar{p}$ events (a) in data and (b) in Monte 
Carlo simulation, with relative branching fractions fixed to data.}
\label{fit}
\end{figure}

     After the above selection, the invariant mass of the proton and 
anti-proton is shown in Fig.~\ref{fit}a, where
clear $\chicz$, $\chico$ and $\chict$ signals can be seen. 
The large peak near the mass of the $\psip$ is due to $\psipto \ppb$ 
with a fake photon 
reconstructed.

The same analysis is performed on a MC sample with 14~M inclusive 
$\psip$ decays generated with Lundcharm~\cite{lundcharm}.
It is found that the remaining 
backgrounds are mainly from $\psipto \piz \ppb$ with 
$\piz \ra 2\gamma$, and they contribute a smooth part to 
the $\ppb$ invariant mass distribution.

\section{\boldmath Fit to the $p \bar{p}$ invariant mass spectrum}

The $p\bar{p}$ mass spectrum of the final selected events with
$p\bar{p}$ mass between 3.26 and 3.64~GeV/$c^2$ is fitted with three
Breit-Wigner resonances smeared by Gaussian mass resolution functions
together with a second order polynomial background using the unbinned
maximum likelihood method. In the fit, the mass resolutions are fixed
to values from Monte Carlo simulation (6.84, 6.59 and 6.17~MeV/$c^2$
for $\chicz$, $\chico$ and $\chict$, respectively), and the widths of
$\chi_{c1}$ and $\chi_{c2}$ are fixed to $0.92$ and $2.08$~MeV/$c^2$
coming from the PDG2002~\cite{pdg}.  The fit, shown in
Fig.~\ref{fit}a, yields a likelihood probability of 79\%, total
numbers of events of $89.5^{+14}_{-13}, 18.2^{+5.5}_{-4.9}$, and
$14.3^{+5.2}_{-4.7}$, and statistical significances of 10.6$\sigma$,
4.6$\sigma$, and 3.7$\sigma$ for $\chi_{c0}$, $\chi_{c1}$ and
$\chi_{c2}$, respectively.  The fitted masses are $3414.3\pm 1.6$,
$3513.0\pm 2.1$, and $3549.1^{+3.2}_{-3.0}$ MeV/$c^2$, respectively,
and agree well with the world averages~\cite{pdg}, and the width of
the $\chicz$ is determined to be $12.8^{+5.6}_{-4.5}$~MeV/$c^2$, in
good agreement with results from other experiments~\cite{pdg}.

The same fit is applied to the Monte Carlo sample, which is about 180 times
larger than the data sample and uses the same relative
branching fractions between the three $\chicJ$ states 
as determined from the data. 
The background fraction is estimated using the $\chicz$ 
mass region for data. The fit yields the efficiencies for each 
channel as
$\varepsilon_{\chi_{c0}}= (27.49\pm0.30)\%$,
$\varepsilon_{\chi_{c1}}= (27.42\pm0.56)\%$, and
$\varepsilon_{\chi_{c2}}= (23.26\pm0.50)\%$,
where the
errors are due to the limited statistics of the Monte Carlo samples.

\section{Systematic Errors}

Systematic errors of the measured branching fractions come from the
efficiencies of photon identification, particle identification,
the kinematic fit, etc.

\subsection{Photon identification}

To investigate the systematic error associated with the
fake photon misidentification, the fake photon multiplicity 
distributions and the energy spectra in
both data and Monte Carlo sample are checked with $\psipto \ppb$
events.  In this channel, it is found that the Monte Carlo simulates a
little more fake photons than data.

The selection of the photon candidate with the largest BSC energy has
an efficiency of $( 92.55\pm 0.52) \%$ for data and $(94.49\pm
0.22)\%$ for Monte Carlo in $\pspto \gamma \chictto \gamma \ppb$,
as determined from the comparison of energy distributions between
the real photons and the fake ones;
the difference is $(2.1\pm 0.6)\%$. For $\chico$ and $\chicz$, the photon
is more energetic, so the efficiency of selecting the largest BSC
energy cluster is higher. The systematic error on the photon
identification is taken as 2.7\% for $\chict$, as well as for $\chicz$
and $\chico$.

\subsection{Photon detection efficiency}

The detection efficiency of low energy photons is studied with
$\jpsito \pp\piz$ events by requiring only one photon in the kinematic
fit and examining the detector response in the direction of emission
of the second photon.  The efficiency from the Monte Carlo simulation
agrees with data within 2\% in the full energy range. This is taken as
the systematic error of the photon detection efficiency.

\subsection{Particle identification}

The systematic error from particle identification is studied using $p$
(or $\bar{p}$) samples from $\psi(2S)\rightarrow \ppjpsi$,
$J/\psi\rightarrow p \bar{p}$ and $\pspto \ppb$.  Since only one
track is required to be identified, the efficiency is very high.  The
correction factors for efficiencies for $\chi_{c0}$, $\chi_{c1}$ and
$\chi_{c2}$ are found to be 1.020$\pm$0.006, 1.025$\pm$0.006 and
1.028$\pm$0.006 respectively.

\subsection{Kinematic fit}

The systematic error associated with the kinematic fit is caused by
differences between the measurements of the momenta and the error
matrices of the track fitting of the charged tracks and the
measurement of the energy and direction of the neutral track for the data
and the simulation sample. The effect is studied for charged tracks and neutral
tracks separately.  The systematic error due to the charged tracks was
checked using $\psipto \ppb$ events, which can be selected easily
without using any kinematic fit.  By comparing the numbers of the
events before and after the kinematic fit, the efficiencies for $\chi^2$
probability $> 1\%$ are measured to be $(85.34\pm 1.64)\%$ for data
and $(88.17\pm 0.56)\%$ for Monte Carlo simulation, respectively.
This results in a correction factor of $(0.968\pm 0.020)\%$ for the
efficiency of this specific channel.

The uncertainty due to the measurement of the neutral track 
parameters is taken from Ref.~\cite{aa} based on a study of 
$\psipto \gamma \chicJ$, $\chicJto \pp\ppb$. 
A systematic error of
4.2\% is quoted for all the channels.

\subsection{Muon veto and BSC energy requirements}
The efficiencies of the muon veto and BSC energy requirements are
studied with $\psipto \ppb$ events. In order to get a clean sample of
$\ppb$ without using these requirements, a four-constraint kinematic
fit to $\ppb$ is done, and the $\chi^2$ probability of the fit is
required to be greater than 1\%. The efficiencies of the muon veto for
both data and Monte Carlo simulation are found to be 100\%, while
those of the BSC energy requirement are $(98.94\pm 0.40)\%$ and
$(99.14\pm 0.18)\%$ for data and Monte Carlo, respectively, resulting
in a difference of $(0.20\pm 0.44)\%$.  This difference together with
the error is taken as the systematic error of the BSC energy
requirement.

\subsection{Angular distribution}

For the radiative decay $\psipto \gamma \chicJ$, the general form of
the angular distribution is $W(\cos\theta) = 1 + A\cos^2\theta$, where
$\theta$ is the angle between the beam direction and the outgoing
photon and $|A|\le 1$~\cite{1203}.  The value of $A$ can be
unambiguously predicted only for spin $J=0$, where $A=1$.  By
comparing the multipole coefficients measured by Crystal
Ball~\cite{cbal} with those used under the assumption of pure $E1$
transition, differences of
3.5\% and 5.5\% are found, which will be taken as systematic errors of
the angular distributions for $J=1$ and $2$ respectively.

\subsection{Breit-Wigner fit}

To determine the fit systematic errors, different background shapes
and fit ranges are used in the fit of the $p\bar{p}$ invariant mass
distribution.  After changing the background shape from a second order
to a first order polynomial and varying the fitting range around the one
used in the fit, the uncertainties due to fit are determined
to be 11.5\%, 8.4\% and 7.3\% for $\chi_{c0}$, $\chi_{c1}$, and
$\chi_{c2}$, respectively.

\subsection{Monte Carlo determined mass resolution}
The systematic error due to the use of the MC determined mass resolution is
studied with $\psi(2S)\rightarrow \ppjpsi$, $J/\psi\rightarrow p
\bar{p}$.  The non-Gaussian tails are found to be ($5.3\pm4.4$)\% and
($5.3\pm1.9$)\% for data and Monte Carlo respectively, which indicates
good agreement between data and Monte Carlo simulation. The
uncertainty of the comparison, 4.8\%, is taken as the systematic error
due to the MC determined mass resolution.

\subsection{Other systematic errors}
The results reported here are based on a data sample corresponding to
a total number of $\psip$ decays, $N_{\psi(2S)}$, of $(14.0 \pm
0.6) \times 10^6$, as determined from inclusive hadronic
events~\cite{pspscan}.  The uncertainty of the number of $\psip$
events, 4\%, is determined from the uncertainty in selecting the
inclusive hadrons.

The difference of MDC tracking efficiencies between data and Monte Carlo 
for $p$ and $\bar{p}$ may cause a systematic error of 1-2\% for each track.
Here 3\% is taken as the systematic error on the overall tracking efficiency.
The trigger efficiency is around 100\% with an uncertainty of 0.5\%, as 
estimated from Bhabha and $e^+ e^- \ra \mu^+ \mu^-$ events.
The systematic errors on the branching fractions used
are obtained from PDG~\cite{pdg} directly.

\subsection{Total systematic error}

Table~\ref{syst} lists the systematic errors from all sources.
Adding all errors in quadrature, the total errors are 17.3\%,
15.4\% and 15.6\% for $\chi_{c0}$, $\chi_{c1}$ and $\chi_{c2}$,
respectively.  The corresponding correction factors~($f$) for the
Monte Carlo simulated efficiencies are
0.987, 0.992, and 0.995.

\begin{table}[htbp]
\caption{Summary of systematic errors in percent. Numbers common for all channels
are only listed once.}
\begin{center}
\begin{tabular}{l|ccc}\hline\hline
 Source&$\chi_{c0}$&$\chi_{c1}$&$\chi_{c2}$\\\hline
 MC statistics &  1.1     &  2.0  & 2.2       \\
 Photon I.D.    & \multicolumn{3}{c}{2.7}\\
  Photon eff. 
                   & \multicolumn{3}{c}{2}  \\
 4C-fit(neutral)   & \multicolumn{3}{c}{$\le$4.2}     \\
 4C-fit(charged)   & \multicolumn{3}{c}{2.0}   \\
 Particle  I.D.      & \multicolumn{3}{c}{0.6} \\
 BW fit            & 11.5 &8.4  & 7.3\\
 Mass resolution   & \multicolumn{3}{c}{4.8}\\
 BSC energy cut    & \multicolumn{3}{c}{0.6}\\
 Angular distr.    & 0 &3.5  & 5.5\\
 Number of $\psip$ & \multicolumn{3}{c}{4}\\
 MDC Tracking      &\multicolumn{3}{c}{3}\\
 Trigger Efficiency&\multicolumn{3}{c}{0.5}\\
$\BR(\psipto \gamma \chicJ)$ \hfil  & 9.2  &  8.3  & 8.8 \\\hline
 Total Systematic error &17.3 &15.4 &15.6\\\hline\hline
\end{tabular}
\end{center}
\label{syst}
\end{table}

\section{Results}

The branching fraction of $\chicJto \ppb$ is calculated using
\begin{displaymath}
   \BR(\chi_{cJ}\rightarrow p\bar{p})=\frac{n^{obs}/(\varepsilon\cdot f)}
{N_{\psp}\cdot B[\pspto \gamma\chi_{cJ}]}.
\end{displaymath}
Using numbers listed in Table.~\ref{br}, one obtains
\begin{eqnarray*}
   \BR(\chiczto \ppb)& =& (27.1^{+4.3}_{-3.9}\pm4.7)\times 10^{-5},\\
   \BR(\chicoto \ppb)& =& (5.7^{+1.7}_{-1.5}\pm0.9)\times  10^{-5},\\
   \BR(\chictto \ppb)& =& (6.5^{+2.4}_{-2.1}\pm1.0)\times  10^{-5},
\end{eqnarray*}
where the first errors are statistical and the second
are systematic.  The measured branching fractions agree with
corresponding world averages within errors~\cite{pdg}.

\begin{table}[htbp]
\caption{Numbers used in branching fraction calculation and the final results.}
\begin{center}
\begin{tabular}{c|ccc}\hline\hline
quantity&$\chi_{c0}$&$\chi_{c1}$&$\chi_{c2}$\\\hline
$n^{obs}$        &$89.5^{+14}_{-13}$ &$18.2^{+5.5}_{-4.9}$&$14.3^{+5.2}_{-4.7}$ \\
$\eff$ (\%)          & 27.49$\pm$0.30&27.42$\pm$0.56 &23.26$\pm$0.50     \\
$f$           &0.987 & 0.992 & 0.995 \\
$N_{\psip}$($10^6$)   &   &14&    \\
$\BR[\psipto \gamma \chicJ]$(\%)&(8.7$\pm$0.8)\%&(8.4$\pm$0.7)\%&(6.8$\pm$0.6)\%\\\hline
$\BR(\chicJto \ppb)$ ($10^{-5}$) &$27.1^{+4.3}_{-3.9}\pm4.7$ 
                                &$5.7^{+1.7}_{-1.5}\pm0.9$
                                &$6.5^{+2.4}_{-2.1}\pm1.0$  \\
$R_{\cal B}$&$1.73\pm0.63$&$4.56\pm2.34$&$5.08\pm3.08$\\\hline
\end{tabular}
\end{center}
\label{br}
\end{table}     

The relative branching fraction of $\chicJto \aab$ to $\chicJto \ppb$ is
found with the following formula:
\begin{displaymath}
  R_{\cal B} = \frac{n^{obs}_{\aab}/[\eff_{\aab} \cdot 
          \BR(\Lambda\ra \pim p)^2]}
                {n^{obs}_{\ppb}/\eff_{\ppb}}.
\end{displaymath}
These results are also shown in Table~\ref{br},
by using the numbers in Table~\ref{br} of this paper 
and those in Table~II of Ref.~\cite{aa},
with the common errors in $\BR(\chicJto \aab)$ and $\BR(\chicJto \ppb)$ 
canceled out.
The measurements confirm the enhancement of 
$\chicJto \aab$ relative to $\chicJto \ppb$, as compared with the 
COM calculation~\cite{wong}.

The actual measured quantities in this analysis is the branching fractions
of $\psipto \gamma \chicJ \ra \gamma \ppb$, with
\begin{eqnarray*}
\BR[\psp \ra \gamma \chicJ \ra \gamma \ppb] & = & 
\BR[\psp \ra \gamma \chicJ]\cdot \BR(\chicJto \ppb)\\
& = & 
\frac{n^{obs}/(\eff\cdot f)}{N_{\psp}},
\end{eqnarray*}
using numbers in Table~\ref{br}, the results are
\begin{eqnarray*}
    \BR[\psip \ra \gamma \chicz \ra \gamma \ppb]&
                =& (23.6^{+3.7}_{-3.4}\pm3.4)\times 10^{-6},\\
    \BR[\psip \ra \gamma \chico \ra \gamma \ppb]&
                =& (4.8^{+1.4}_{-1.3}\pm0.6 )\times 10^{-6},\\
    \BR[\psip \ra \gamma \chict \ra \gamma \ppb]&
                =& (4.4^{+1.6}_{-1.4}\pm0.6 )\times 10^{-6}.
\end{eqnarray*}

\section{Summary}

Decays $\chicJ \ra \ppb$ are observed using the BESII sample of 14
million $\psip$ events, and the corresponding branching fractions are
determined. The measured values agree with previous experiments within
errors~\cite{e8351,e8352,e760,besi}.
The measurements confirm the enhancement of 
$\chicJto \aab$ relative to $\chicJto \ppb$, as compared with the 
COM calculation~\cite{wong}.

\acknowledgments

   The BES collaboration thanks the staff of the BEPC for their 
hard efforts. This work is supported in part by the National 
Natural Science Foundation of China under contracts 
Nos. 19991480, 10225524, 10225525, the Chinese Academy
of Sciences under contract No. KJ 95T-03, the 100 Talents 
Program of CAS under Contract Nos. U-11, U-24, U-25, and 
the Knowledge Innovation Project of CAS under Contract 
Nos. U-602, U-34(IHEP); by the National Natural Science
Foundation of China under Contract No. 10175060 (USTC); 
and by the U.S. Department of Energy under Contract 
No. DE-FG03-94ER40833 (U Hawaii).

\end{document}